\documentclass{article}

     \PassOptionsToPackage{numbers, compress}{natbib}
  \usepackage[preprint]{neurips_2026}

\usepackage[utf8]{inputenc} 
\usepackage[T1]{fontenc}    
\usepackage{hyperref}       
\usepackage{url}            
\usepackage{booktabs}       
\usepackage{amsfonts}       
\usepackage{nicefrac}       
\usepackage{microtype}      
\usepackage{xcolor}         

\usepackage{amsmath,amssymb,amsfonts,amsthm,bm}
\usepackage{graphicx}
\usepackage{booktabs}
\usepackage{multirow}
\usepackage{array}
\usepackage{xcolor}
\usepackage{caption}
\usepackage{subcaption}
\usepackage{algorithm}
\usepackage{algpseudocode}

\graphicspath{{../}{NeurIPS_plots/}}

\newtheorem{theorem}{Theorem}[section]

\title{High-Dimensional Clustering via Nearest-Neighbor Asymmetry}

\author{%
  Hao Chen and Xiancheng Lin \\
  Department of Statistics\\
  University of California, Davis\\
  Davis, CA 95616 \\
  \texttt{hxchen@ucdavis.edu; xclin@ucdavis.edu} \\
}

\begin{document}

\maketitle

\begin{abstract}
High-dimensional clustering often relies on geometric or local-similarity structure, but the dominant separation between groups may not always be location-based. Differences in dispersion can create asymmetric local-neighborhood patterns: points from a more dispersed component may be closer to points in a more concentrated component than to points from their own component. We turn this high-dimensional phenomenon into a clustering principle. The proposed method, NAC (Nearest-neighbor Asymmetry Clustering), constructs a directed $k$-nearest-neighbor graph and evaluates candidate partitions using two permutation-standardized statistics: a weighted within-edge statistic that captures overall within-cluster enrichment and a contrast statistic that captures asymmetric separation. The resulting objective combines these two standardized signals, allowing the method to adapt to different separation regimes without specifying a mixture model or a low-dimensional representation. We provide a population-level analysis showing how the two statistics target complementary nearest-neighbor patterns. Simulation studies across mean, scale, and combined location-scale differences show that NAC is competitive under location separation and especially effective when nearest-neighbor asymmetry is present; gene-expression applications further illustrate its usefulness in small-sample, high-dimensional clustering.
\end{abstract}

\section{Introduction}
\label{sec:intro}

Clustering high-dimensional data is a central task in modern machine learning.
A common motivation for clustering is to obtain groups that can be interpreted,
summarized, or analyzed separately, often because observations in different
groups arise from different underlying data-generating mechanisms, such as
different distributions. Geometric closeness and local similarity are powerful
ways to capture such structure, and many popular methods are highly effective
when clusters are separated by location or form locally coherent geometric
regions. Representative examples include centroid-based methods such as
$k$-means, spectral and other graph partitioning methods, and embedding-based
approaches followed by clustering
\citep{mcqueen1967some,hastie2009elements,shi2000normalized,ng2001spectral,
von2007tutorial,maaten2008visualizing}.

A distinct high-dimensional challenge arises when the difference between groups
appears mainly through scale or variability. Such scale differences can distort
local neighborhoods, so geometric closeness may no longer reflect the desired
grouping. This creates a challenge for clustering methods that rely on
geometric or similarity-based structure: when separation is scale-driven rather
than location-driven, many common methods can miss the relevant signal,
depending on how they encode geometry or local similarity.

A simple example illustrates the issue. Consider two samples of size 50 from
$\mathcal N_d(0,\Sigma)$ and $\mathcal N_d(a\mathbf 1,b\Sigma)$, where $d=200$
and $\Sigma_{ij}=0.1^{|i-j|}$. Table~\ref{tab:motivation} reports mis-clustering rates from an illustrative
simulation run for several representative existing methods: spectral clustering as
implemented by \texttt{specc} in the \texttt{R} package \texttt{kernlab}
(Specc) \citep{karatzoglou2004kernlab}; regularized spectral clustering based
on adjacency and Laplacian matrices (RSpec(A) and RSpec(L))
\citep{rohe2011spectral}; IF-PCA, a high-dimensional
clustering method that performs PCA after selecting influential features \citep{jin2016influential}; and $t$-SNE
followed by $k$-means ($t$-SNE) \citep{maaten2008visualizing}. Under mean separation $(a,b)=(0.5,1)$, some methods perform well, with Specc
achieving a particularly low mis-clustering rate. Under scale separation
$(a,b)=(0,1.4)$, however, the baseline methods have substantially higher
mis-clustering rates. The final column reports NAC, the method developed in this paper, which achieves
a substantially lower mis-clustering rate, consistent with its use of the
asymmetric nearest-neighbor pattern induced by scale separation.

\begin{table}[t]
\centering
\caption{Mis-clustering rates under mean and scale differences. Best in each row is bolded.}
\label{tab:motivation}
\begin{tabular}{lcccccc}
\toprule
Setting & Specc & RSpec(A) & RSpec(L) & IF-PCA & t-SNE & NAC \\
\midrule
Mean differs 
& \textbf{0.008} & 0.291 & 0.175 & 0.293 & 0.108 & \textbf{0.008} \\
Scale differs 
& 0.347 & 0.416 & 0.361 & 0.429 & 0.413 & \textbf{0.146} \\
\bottomrule
\end{tabular}
\end{table}

This contrast suggests that useful cluster structure may be encoded not by
uniformly strong within-cluster connectivity, but by an asymmetric
nearest-neighbor pattern: points in the less dispersed cluster tend to choose
neighbors from their own cluster, whereas points in the more dispersed cluster
may often choose neighbors from the less dispersed cluster.

We turn this observation into a clustering principle. NAC, short for
nearest-neighbor asymmetry clustering, evaluates a candidate binary partition on
a directed $k$-nearest-neighbor graph using two permutation-standardized graph
statistics. The first measures overall within-cluster edge enrichment and is
effective for location-like separation. The second measures the difference
between the two within-cluster edge counts and is designed to capture
asymmetric scale-like separation. NAC takes the maximum of the two standardized
signals, allowing either symmetric enrichment or asymmetric nearest-neighbor
imbalance to support a partition.

The resulting method is nonparametric and does not require specifying a mixture
model, selecting a low-dimensional representation, or assuming that all clusters
are internally dense in the same way. Instead, it searches for a labeling whose
nearest-neighbor structure is most unlikely under random relabeling. This makes
the method useful in high-dimensional settings where the type of separation is
unknown and may involve location, scale, or other distributional differences.

The main contribution is to show that asymmetric nearest-neighbor structure,
which may appear to be a failure mode of similarity-based clustering, can be
used as a clustering signal in high dimensions. The sections below formalize
this principle (Section~\ref{sec:nn-asymmetry}), develop the NAC criterion
(Section~\ref{sec:method}), describe the optimization procedure
(Section~\ref{sec:optimization}), and evaluate NAC across synthetic and
gene-expression examples
(Sections~\ref{sec:experiments}--\ref{sec:applications}).


\section{Nearest-neighbor asymmetry in high dimensions}
\label{sec:nn-asymmetry}

We first describe the nearest-neighbor pattern that motivates the proposed
criterion. Let $G_k$ be the directed $k$-nearest-neighbor graph constructed from
the pooled observations, with an edge $(i,j)$ if observation $j$ is among the
$k$ nearest neighbors of observation $i$. For a candidate binary partition $x$,
let $R_{1,k}(x)$ and $R_{2,k}(x)$ denote the numbers of directed edges with both
endpoints in the first and second clusters, respectively. For a candidate
partition with cluster sizes $m$ and $n$, define the empirical
within-cluster neighbor proportions
\[
p_{11,k}(x)=\frac{R_{1,k}(x)}{km},\qquad
p_{22,k}(x)=\frac{R_{2,k}(x)}{kn}.
\]
Under random relabeling of the $N=m+n$ observations, with $m$ assigned to
cluster 1 and $n$ assigned to cluster 2, the expected within-cluster neighbor
proportions are
\[
p_{11}^{(0)}=\frac{m-1}{N-1}
\quad \text{and} \quad
p_{22}^{(0)}=\frac{n-1}{N-1}.
\]

\begin{figure}[t]
\centering
\includegraphics[width=.32\textwidth]{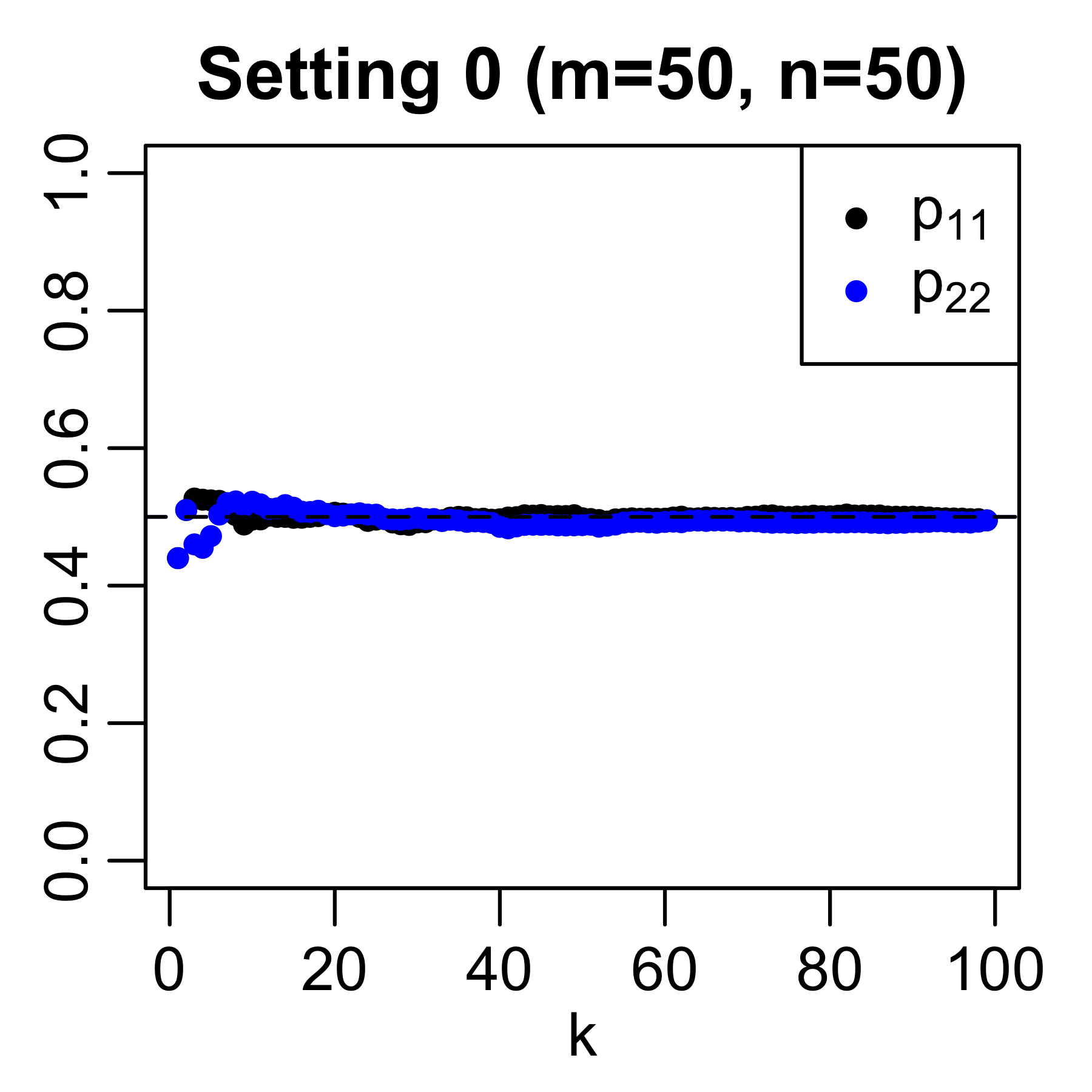}
\includegraphics[width=.32\textwidth]{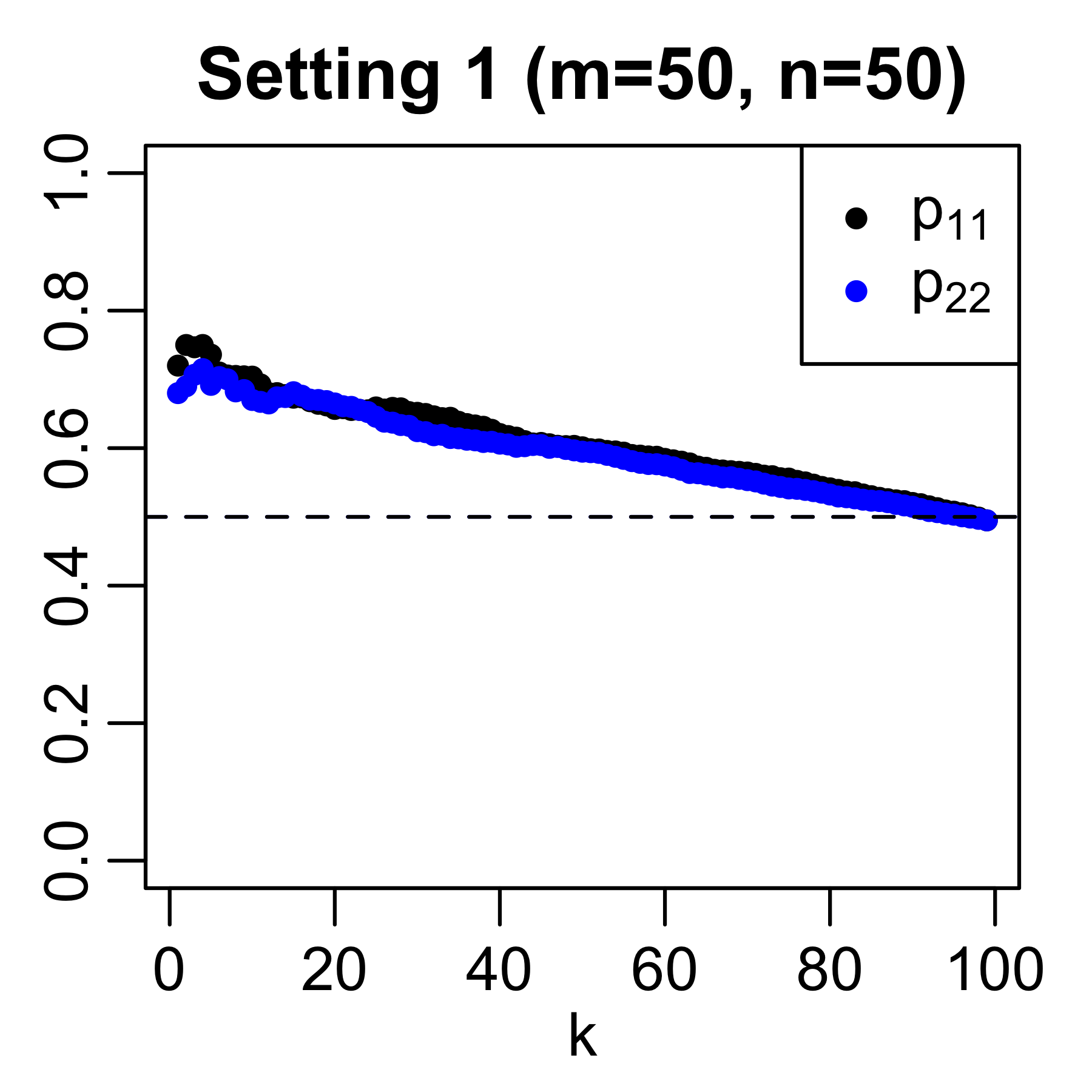}
\includegraphics[width=.32\textwidth]{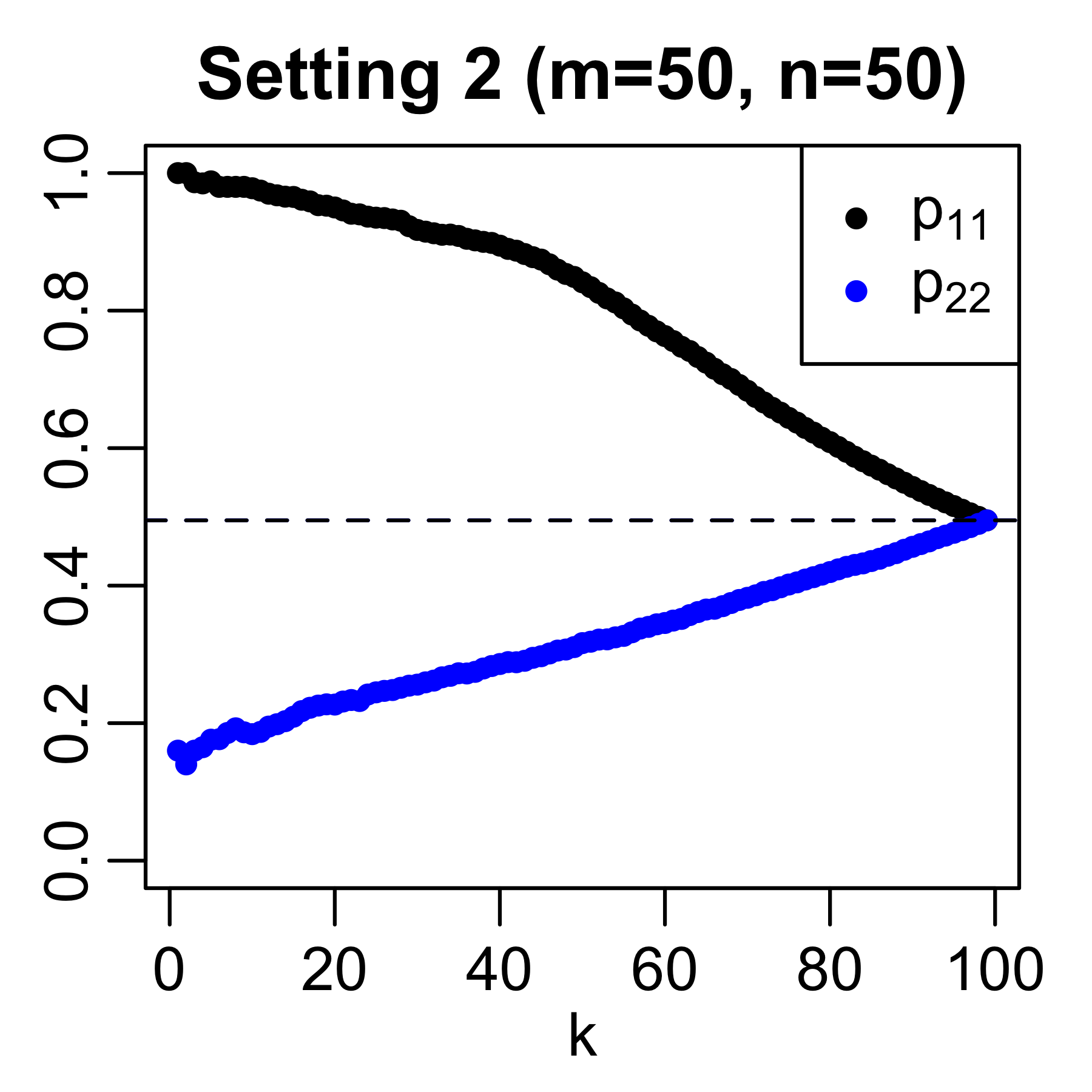}
\includegraphics[width=.32\textwidth]{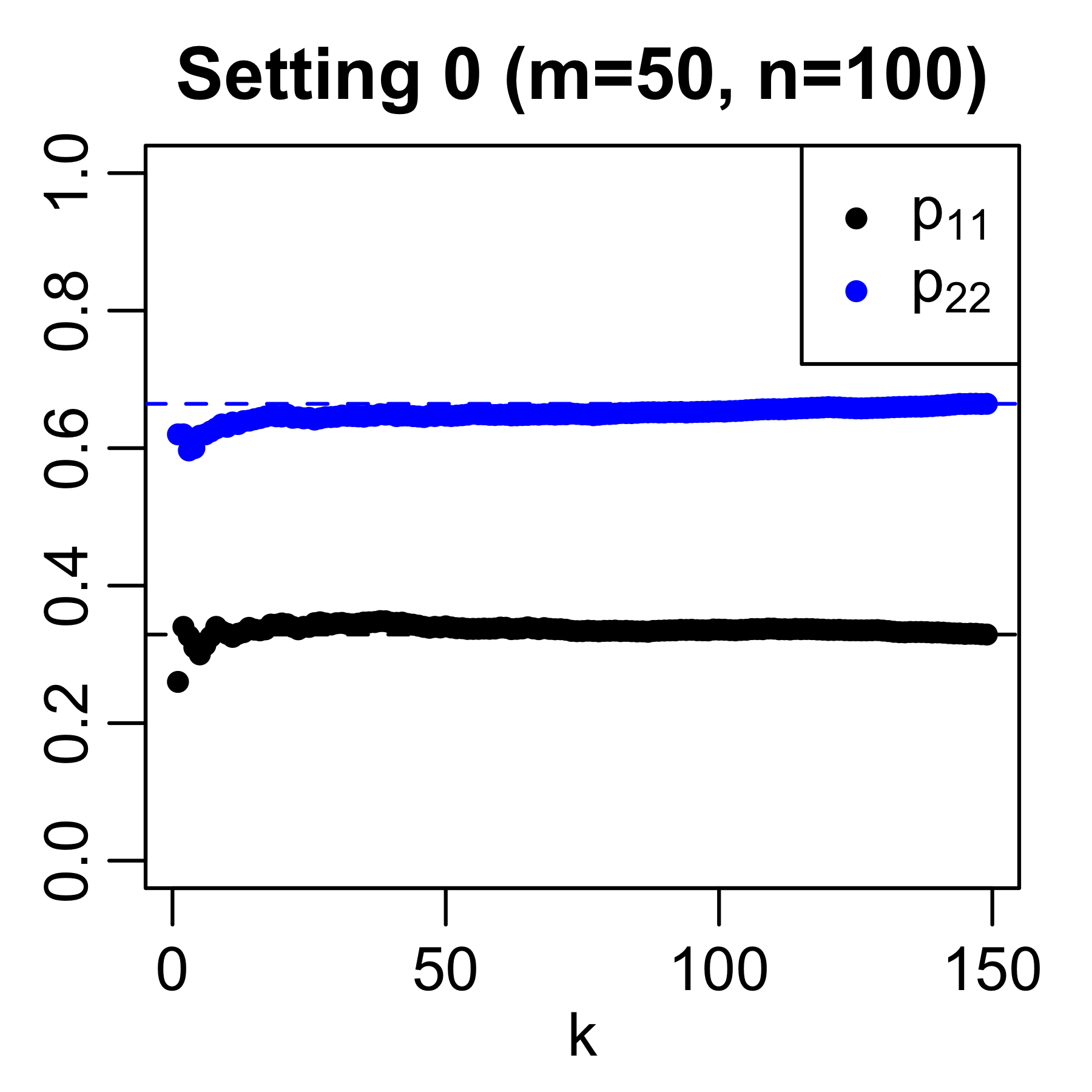}
\includegraphics[width=.32\textwidth]{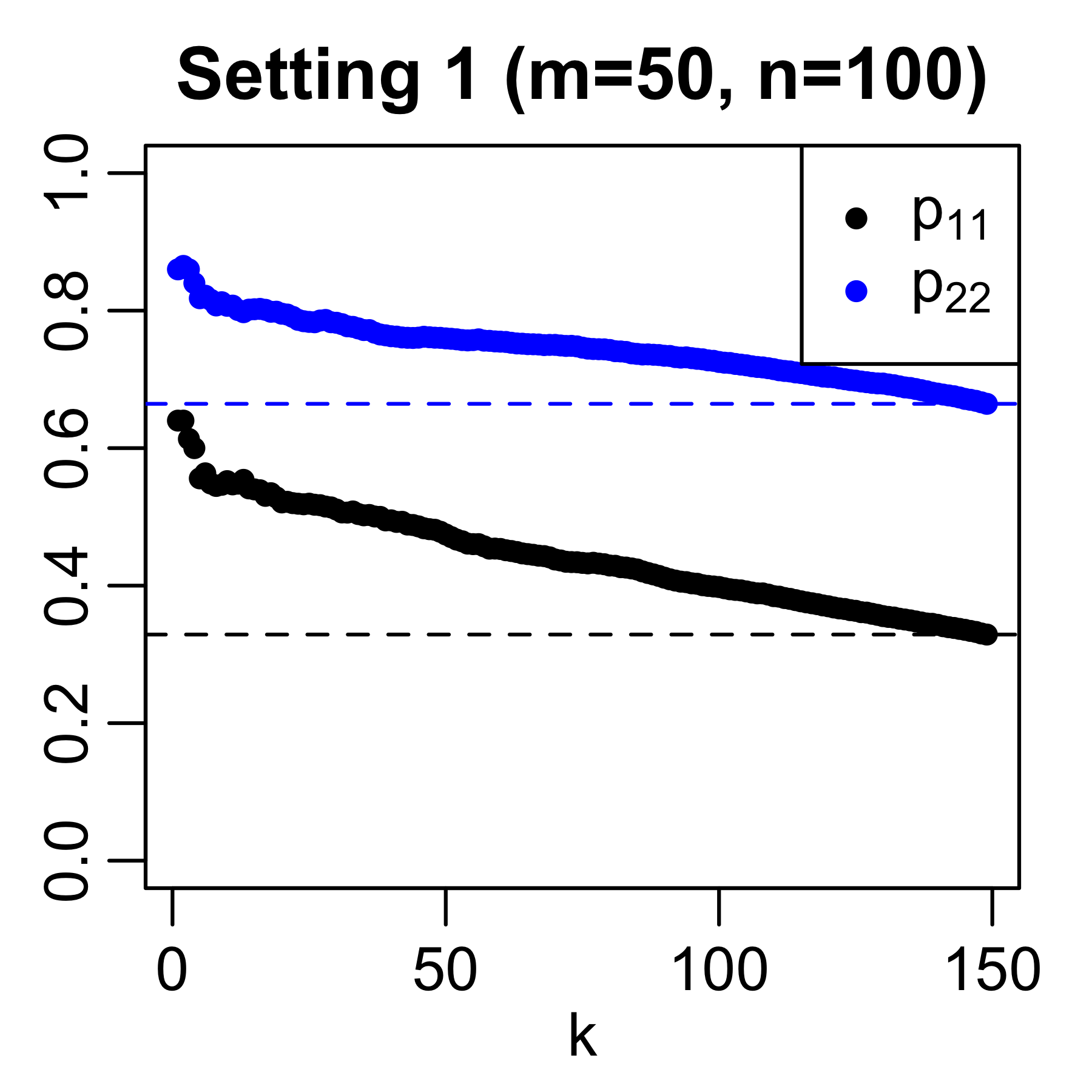}
\includegraphics[width=.32\textwidth]{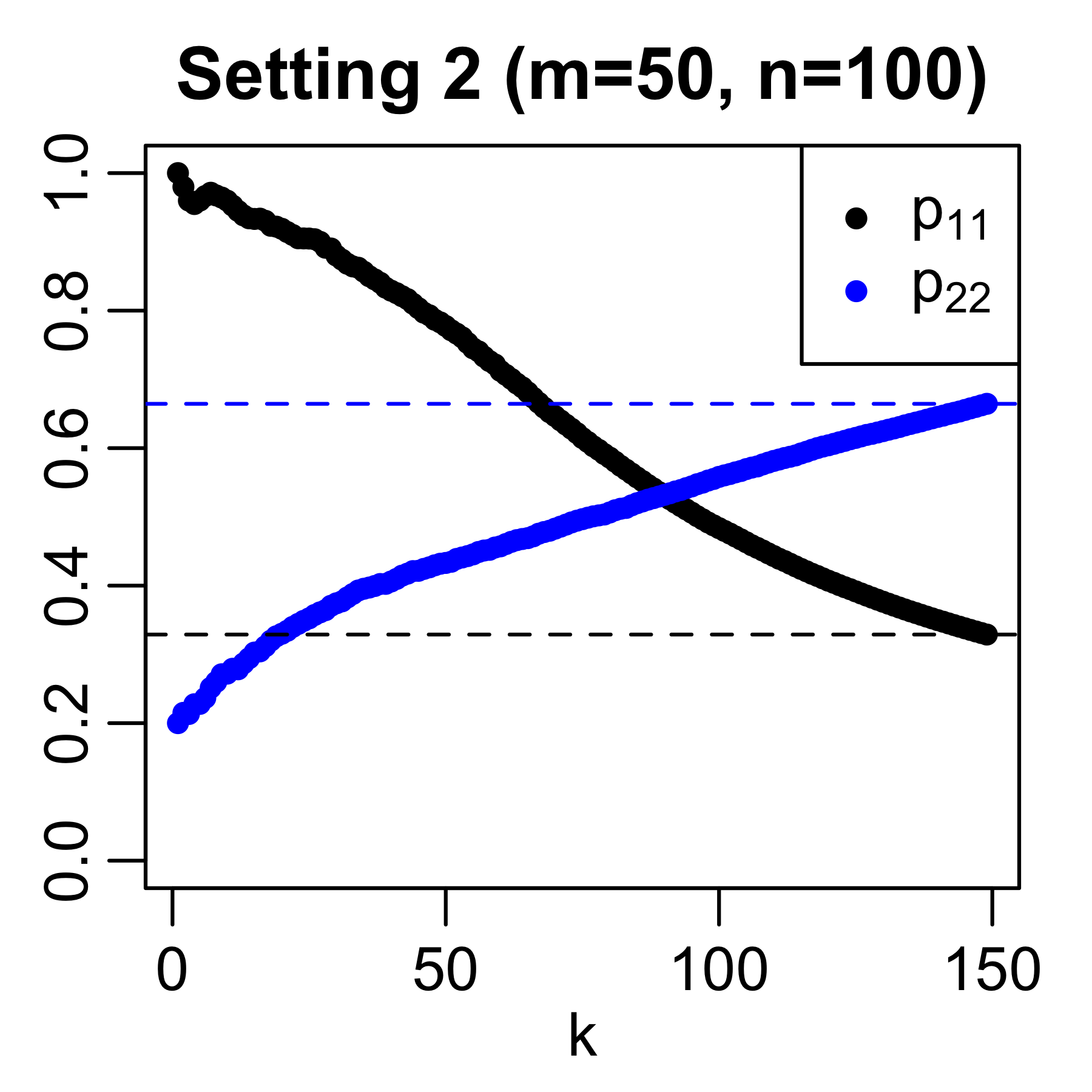}
\caption{Within-cluster nearest-neighbor proportions under the three Gaussian
settings defined in the text. The top row uses balanced sample sizes
$(m,n)=(50,50)$ and the bottom row uses unbalanced sample sizes
$(m,n)=(50,100)$. Location separation tends to increase within-cluster neighbor
proportions for both groups, whereas scale differences can produce an
asymmetric pattern.}
\label{fig:pii-main}
\end{figure}

To see how location and scale differences affect these proportions relative to
their random-labeling baselines, consider two clusters generated from
$\mathcal N_d(0,\Sigma)$ and $\mathcal N_d(a\mathbf 1,b\Sigma)$, where
$d=200$ and $\Sigma_{ij}=0.1^{|i-j|}$. We consider three representative
settings: no distributional difference (Setting 0, $(a,b)=(0,1)$), location
difference with comparable scales (Setting 1, $(a,b)=(0.3,1)$), and combined
location-scale differences (Setting 2, $(a,b)=(0.3,1.3)$).
Figure~\ref{fig:pii-main} plots $p_{11,k}$ and $p_{22,k}$ across graph
sizes $k$, together with their random-labeling baselines.

Figure~\ref{fig:pii-main} shows that location and scale differences lead to
qualitatively different nearest-neighbor patterns. In Setting 0, the empirical
proportions remain close to their random-labeling baselines. In Setting 1, where
the clusters differ in location but have comparable scales, both $p_{11,k}$ and
$p_{22,k}$ tend to exceed their baselines: points from each cluster
preferentially choose neighbors from the same cluster. In Setting 2, where a
scale difference is also present, the pattern becomes asymmetric. The less
dispersed cluster can retain an above-baseline within-cluster neighbor
proportion, while the more dispersed cluster can fall below its baseline. Thus
the informative departure from random labeling is not always a uniformly high
level of within-cluster connectivity; it may instead be an imbalance between
the two within-cluster neighbor proportions.

This phenomenon is particularly pronounced in high dimensions because small
relative differences in scale can have large effects on pairwise distances. For
a more dispersed component, distances to other points from the same component
can grow enough that points in a less dispersed component become competitive
nearest neighbors. Thus a scale difference does not simply make one group
``larger''; it changes the direction of local nearest-neighbor edges. This is
the sense in which a curse-of-dimensionality pattern becomes useful here: it
creates a stable imbalance in the within-cluster neighbor proportions rather
than only making all distances less informative.

This distinction has direct consequences for clustering criteria. A criterion
based only on the total within-edge count, or equivalently on uniformly strong
within-cluster connectivity, is well matched to symmetric enrichment but can
miss scale-driven separation. In that regime, adding the two within-cluster
counts can average away the most informative feature of the graph: one group is
less dispersed and locally coherent, while the other is more dispersed.

To see this more explicitly, write the deviations from the random-labeling
baselines as
\[
\eta_1(x)=p_{11,k}(x)-p_{11}^{(0)},\qquad
\eta_2(x)=p_{22,k}(x)-p_{22}^{(0)}.
\]
A total within-edge criterion adds the two cluster-specific deviations, after
accounting for cluster sizes, into a single overall enrichment measure. This is appropriate under symmetric enrichment,
where both deviations are positive. Under the asymmetric pattern described
above, however, the two deviations have opposite signs and can offset each
other, so the sum may be small even when the contrast
$|\eta_1(x)-\eta_2(x)|$ is large.

The next section turns this observation into a clustering objective. The key is
to form one statistic for overall within-cluster enrichment and another for
within-cluster imbalance, then standardize both under random relabeling so that
candidate partitions with different sizes and graph structures can be compared
on a common scale.

\section{NAC: A nearest-neighbor asymmetry clustering criterion}
\label{sec:method}

Let $X_1,\ldots,X_N$ be the pooled observations and let
$G_k=(V,E_k)$ be the directed $k$-nearest-neighbor graph. A candidate binary
partition is encoded by $x=(x_1,\ldots,x_N)\in\{0,1\}^N$, where $x_i=1$ assigns
observation $i$ to the first cluster and $x_i=0$ assigns it to the second. We
write the criterion conditional on the candidate cluster size, with
$\sum_i x_i=m$ and $n=N-m$; during optimization, $m$ is determined by the
candidate partition, and the criterion is evaluated using the corresponding
values of $m$ and $n$.

For a candidate partition $x$, the within-cluster edge counts can be written as
\[
R_{1,k}(x)=\sum_{(i,j)\in E_k}x_i x_j,\qquad
R_{2,k}(x)=\sum_{(i,j)\in E_k}(1-x_i)(1-x_j).
\]
These count directed $k$-nearest-neighbor edges within the first and second
clusters, respectively. When the graph size is fixed, we suppress the subscript
$k$ and write $R_1(x)$ and $R_2(x)$. If the labels were observed, these would be
directed within-sample edge counts used in nearest-neighbor graph two-sample
tests \citep{schilling1986multivariate,henze1988multivariate}. In clustering,
the labels are unknown, so the same graph summaries must be evaluated over
candidate partitions.

To capture the two patterns identified in Section~\ref{sec:nn-asymmetry}, NAC
uses two summaries of the within-cluster edge counts. The first is a weighted
within-edge statistic,
\[
R_w(x)=\frac{n-1}{N-2}R_1(x)+\frac{m-1}{N-2}R_2(x).
\]
The weights adjust for the candidate cluster sizes and match the weighted
edge-count statistic used in graph-based two-sample testing \citep{chen2017new}.
This statistic is designed for the symmetric-enrichment pattern: it becomes
large when the candidate partition has more within-cluster nearest-neighbor
edges in aggregate than expected under random labeling. The second summary is
the difference statistic
\[
R_d(x)=R_1(x)-R_2(x).
\]
Unlike a total within-edge count, this contrast does not average away the two
cluster-specific signals. It is designed for the asymmetric-enrichment pattern:
one cluster may have an elevated within-cluster edge count, while the other may
not. The sign of $R_d$ depends on the arbitrary labeling of the two clusters, so
we use its absolute standardized deviation below.

Because candidate partitions can have different sizes and because the graph
itself induces dependence among edge counts, raw values of $R_w$ and $R_d$ are
not directly comparable. We therefore standardize them under a permutation null,
following the random-labeling principle used in graph-based two-sample testing
\citep{friedman1979multivariate,schilling1986multivariate,
henze1988multivariate,rosenbaum2005exact,chen2017new}. Under this null, the
graph $G_k$ is fixed and the labels are randomly permuted with $m$ observations
assigned to cluster 1 and $n=N-m$ assigned to cluster 2.

Let $\mathbb E_0$ and $\mathrm{Var}_0$ denote expectation and variance under this
random-labeling distribution. For the two statistics above,
\[
\mu_w=\mathbb E_0R_w
=\frac{(m-1)(n-1)}{(N-1)(N-2)}\,kN,
\qquad
\mu_d=\mathbb E_0R_d
=k(m-n).
\]
The variances depend on the dependence structure of the observed directed graph.
Let
\[
q_{1,k}=\sum_{(i,j)\in E_k}\mathbf 1\{(j,i)\in E_k\},
\qquad
q_{2,k}=\sum_{i=1}^N\sum_{j\ne \ell}\mathbf 1\{(j,i),(\ell,i)\in E_k\}.
\]
Here $q_{1,k}$ counts mutual directed nearest-neighbor pairs, and $q_{2,k}$
counts ordered pairs of directed edges sharing the same terminal node. The
permutation null variances are
\[
\begin{aligned}
\sigma_w^2
&=\mathrm{Var}_0(R_w)=\frac{mn(m-1)(n-1)}{N(N-1)(N-2)(N-3)}
\left\{
kN+q_{1,k}
-\frac{q_{2,k}+kN-k^2N}{N-2}
-\frac{2k^2N}{N-1}
\right\},\\
\sigma_d^2
&=\mathrm{Var}_0(R_d)=\frac{mn}{N(N-1)}
\left(q_{2,k}+kN-k^2N\right).
\end{aligned}
\]
Define the standardized statistics
\[
Z_w(x)=\frac{R_w(x)-\mu_w}{\sigma_w},\qquad
Z_d(x)=\frac{|R_d(x)-\mu_d|}{\sigma_d}.
\]
The statistic $Z_w$ measures standardized within-neighbor enrichment, while
$Z_d$ measures standardized asymmetry between the two within-cluster edge
counts. The absolute value in $Z_d$ makes the statistic invariant to switching
the two cluster labels.

The expressions for $\mu_w$, $\mu_d$, $\sigma_w^2$, and $\sigma_d^2$ follow from the same combinatorial arguments used for
graph-based permutation statistics in two-sample testing
\citep{schilling1986multivariate,henze1988multivariate,chen2017new} and
change-point detection \citep{liu2022fast}. They also show why the conditional
standardization is central to the method. Two $k$-nearest-neighbor graphs with
the same $N$ and $k$ can have different null variability if one contains many
mutual nearest-neighbor pairs or many shared nearest-neighbor targets. NAC
therefore standardizes each candidate partition relative to the actual graph
geometry rather than treating all $k$-nearest-neighbor graphs as equivalent.

NAC combines the two standardized signals through
\[
M_{\kappa,k}(x)=\max\{Z_w(x),\kappa Z_d(x)\},
\]
where $\kappa>0$ balances the symmetric and asymmetric components. For a fixed graph size $k$, the estimated clustering is
\[
\hat x_k\in\arg\max_x M_{\kappa,k}(x),
\]
where the maximization is over nontrivial binary partitions for which both
clusters are nonempty and the standardized statistics are well defined. The
final clustering is interpreted up to label switching.

This objective differs from standard graph-cut criteria such as MinCut
\citep{stoer1997simple}, RatioCut
\citep{leighton1988approximate,chan1993spectral}, and normalized cut
\citep{shi2000normalized}. Cut-based methods are based on between-cluster
edges, or equivalently on total within-cluster connectivity, and therefore
favor partitions in which both clusters are internally well connected. The
statistic $Z_w$ has a similar spirit after permutation standardization. The
statistic $Z_d$, however, retains the imbalance between the two within-cluster
edge counts instead of averaging it away. This distinction is essential in the
scale-separated regime, where one cluster may be less dispersed and locally
coherent while the other is more dispersed.

The maximum in $M_{\kappa,k}$ lets either type of evidence support a partition.
This is useful because the relevant signal may come primarily from $Z_w$ under
symmetric enrichment or from $Z_d$ under asymmetric enrichment. Thus NAC can
adapt to the graph pattern without requiring the user to specify in advance
whether the separation is location-like or scale-like.

The following population-level result formalizes the patterns illustrated in
Figure~\ref{fig:pii-main}. These patterns correspond to symmetric enrichment,
where both clusters have above-baseline within-cluster neighbor proportions, and
asymmetric enrichment, where one cluster is above baseline and the other is
below. We distinguish the following cases:
\[
\begin{array}{ll}
\textbf{Pattern 1: symmetric enrichment,} &
p_{11,k}(x^*)>p_{11}^{(0)}, \qquad
p_{22,k}(x^*)>p_{22}^{(0)},\\[3pt]
\textbf{Pattern 2: asymmetric enrichment,} &
p_{11,k}(x^*)>p_{11}^{(0)}, \qquad
p_{22,k}(x^*)<p_{22}^{(0)},\\[3pt]
\textbf{Pattern 3: asymmetric enrichment,} &
p_{11,k}(x^*)<p_{11}^{(0)}, \qquad
p_{22,k}(x^*)>p_{22}^{(0)}.
\end{array}
\]

\begin{theorem}[Population motivation]
\label{thm:criterion}
Fix a graph scale $k$. At the population level, with within-label edge counts
replaced by their conditional expectations under the limiting directed-neighbor
probabilities, Pattern 1 implies that $Z_w$ is maximized by the true partition,
up to label switching. Patterns 2 and 3 imply that $Z_d$ is maximized by the
true partition, up to label switching.
\end{theorem}

The proof is given in Appendix~\ref{app:proof}. The theorem identifies the
population-level signal targeted by the two components: if the true partition
creates symmetric enrichment, $Z_w$ is the relevant diagnostic; if it creates
asymmetric enrichment, $Z_d$ is the relevant diagnostic. The optimization
procedure in the next section gives a practical finite-sample search strategy.

\section{Optimization and tuning}
\label{sec:optimization}

The objective $M_{\kappa,k}(x)$ is combinatorial, so exact maximization over
all binary partitions is infeasible except for very small samples. We use a
multi-start local search for each fixed graph size $k$. The search is
initialized from several candidate partitions. Starting from each
initialization, we iteratively update the labels to increase $M_{\kappa,k}(x)$.
Among all starts, we retain the partition with the largest objective value.

A single update can be evaluated efficiently. Moving one observation changes the
within-cluster status only for directed edges leaving that observation and for
directed edges whose terminal node is that observation. Thus the objective can
be updated without recomputing all edge counts from scratch. In practice, the absolute value in $Z_d$ makes the criterion invariant to label switching,
so the result does not depend on which cluster is called cluster 1. The
multi-start strategy is important because the objective is nonconvex and a
single initialization can converge to a suboptimal local maximum.

\begin{figure}[!b]
\centering
\includegraphics[width=.31\textwidth]{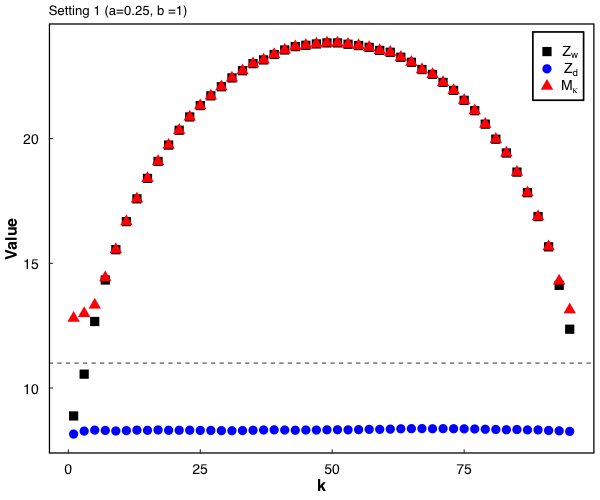}
\includegraphics[width=.31\textwidth]{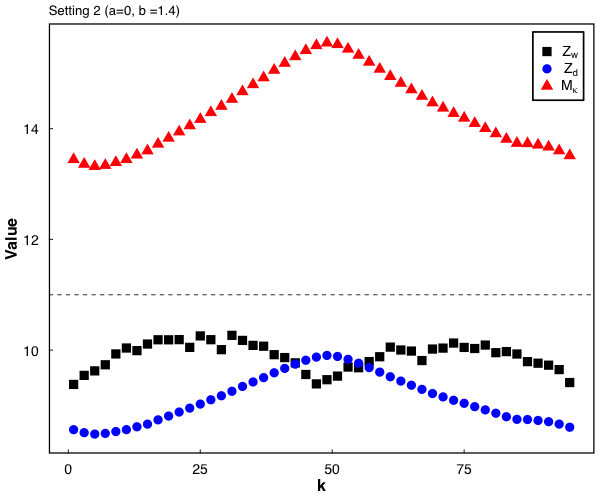}
\includegraphics[width=.31\textwidth]{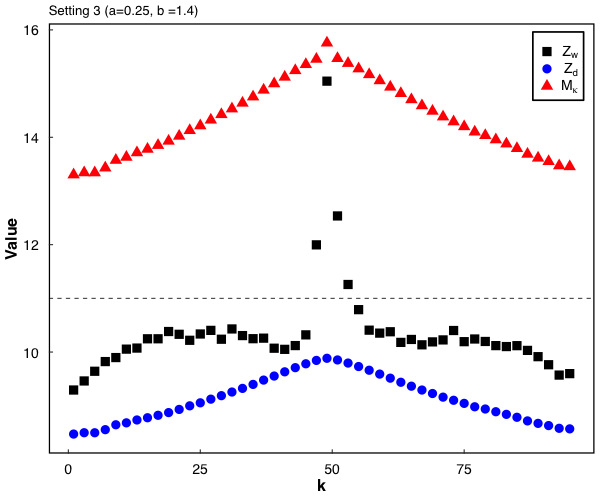}
\includegraphics[width=0.322\textwidth]{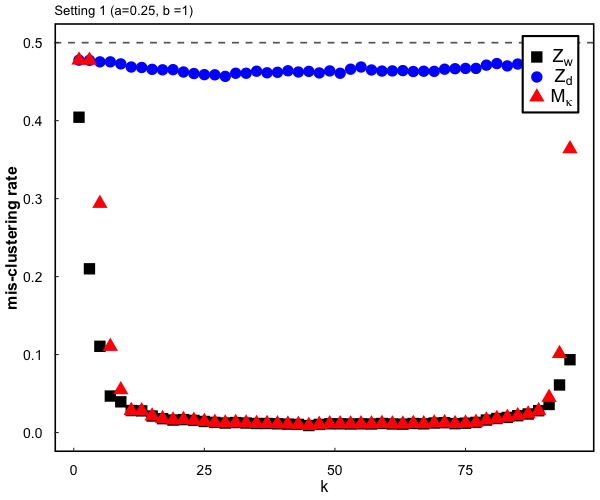}
\includegraphics[width=0.322\textwidth]{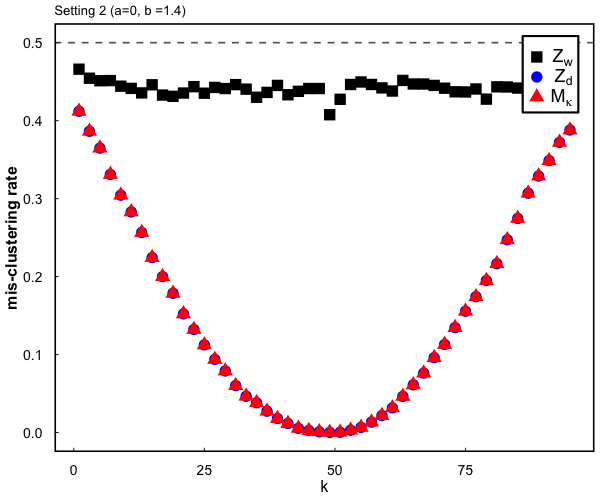}
\includegraphics[width=0.322\textwidth]{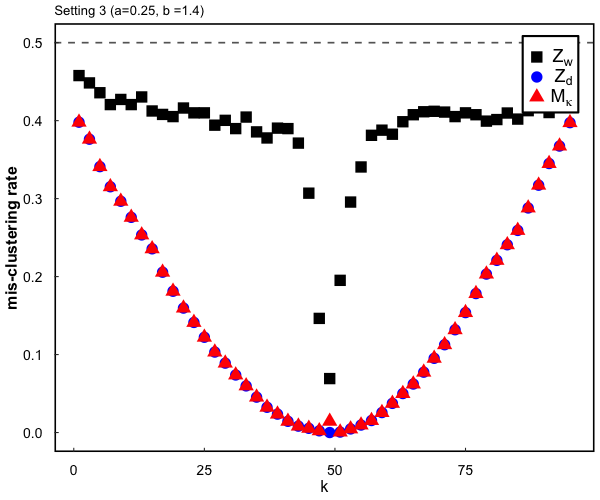}
\caption{Effect of graph size $k$ in representative Gaussian settings with
$m=n=50$. The top row shows the optimized values of $Z_w$, $Z_d$, and
$M_{\kappa,k}$; the bottom row shows the corresponding mis-clustering rates.
\textbf{Large values of $M_{\kappa,k}$ tend to occur near graph sizes with low
mis-clustering rates.} The useful component and the best graph size vary across
location-like, scale-like, and combined alternatives, motivating the maximum
criterion and graph-size selection over a grid of $k$ values.}
\label{fig:k-diagnostic}
\end{figure}

The graph size $k$ controls the local scale of the nearest-neighbor graph.
Small values of $k$ can make the edge counts noisy, while large values can wash
out local structure. For this diagnostic and for all experiments below, we use
the fixed default $\kappa=1.55$; its calibration is described at the end of this
section. Figure~\ref{fig:k-diagnostic} illustrates the role of $k$ in
representative Gaussian settings. The top row shows the optimized values of
$Z_w$, $Z_d$, and $M_{\kappa,k}$, while the bottom row shows the corresponding
mis-clustering rates. Across these settings, large values of $M_{\kappa,k}$
tend to occur near graph sizes with low mis-clustering rates, suggesting that
the criterion can serve as a useful label-free guide to informative graph
scales.

The useful component and the best graph size vary across location-like,
scale-like, and combined alternatives, which motivates evaluating a grid of
$k$ values rather than fixing a single graph size. For each $k$ in the grid,
the local search returns a candidate partition and its optimized objective
value. Since $M_{\kappa,k}$ is fully observable from the data, we select the
graph size that gives the largest optimized value of $M_{\kappa,k}$ and use the
corresponding partition as the final clustering, up to label switching. An
optional faster graph-size search based on the separate component objectives is
provided in Appendix~\ref{app:optimization}.

The tuning parameter $\kappa$ balances the symmetric statistic $Z_w$ and the
asymmetric statistic $Z_d$. Small values of $\kappa$ emphasize conventional
within-neighbor enrichment, whereas large values emphasize asymmetric
nearest-neighbor structure. Although $Z_w$ and $Z_d$ are standardized before
optimization, their optimized values are not identically distributed. In the
diagnostic above and in all experiments below, we use a fixed default
$\kappa=1.55$, chosen from representative location and scale settings. This
single value is held fixed throughout, avoiding any separate tuning of
$\kappa$ for individual simulation settings or real datasets. Calibration
diagnostics are provided in Appendix~\ref{app:kappa}.

\section{Experiments}
\label{sec:experiments}

We evaluate NAC against representative clustering baselines from several common
families: spectral clustering (Spectral) \citep{hastie2009elements}, spectral clustering implemented by
\texttt{specc} (Specc) \citep{karatzoglou2004kernlab}, regularized spectral clustering based on adjacency and
Laplacian matrices (RSpec(A) and RSpec(L)) \citep{rohe2011spectral}, IF-PCA \citep{jin2016influential}, and $t$-SNE followed by
$k$-means \citep{maaten2008visualizing}. Details on the \texttt{specc} kernel-width setting for high-dimensional data are
provided in Appendix~\ref{app:specc-kpar}. Performance is measured by the
mis-clustering rate after optimal label switching. Unless otherwise stated,
results are averaged over repeated simulations with two clusters of size 50.

\subsection{Gaussian mixtures}
\label{sec:gaussian}

\begin{figure}[!b]
\centering
\includegraphics[width=.32\textwidth]{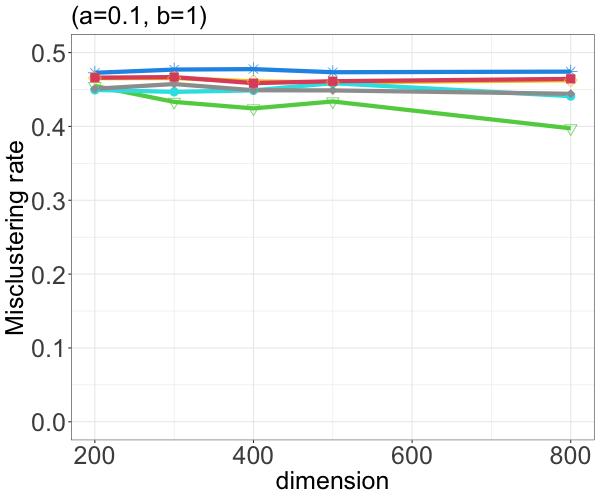}
\includegraphics[width=.32\textwidth]{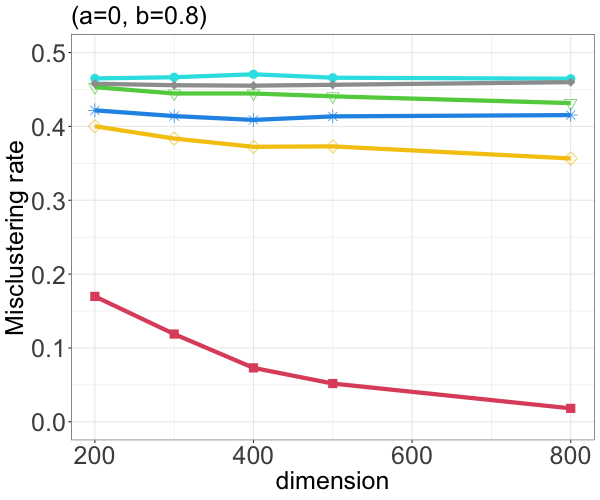}
\includegraphics[width=.32\textwidth]{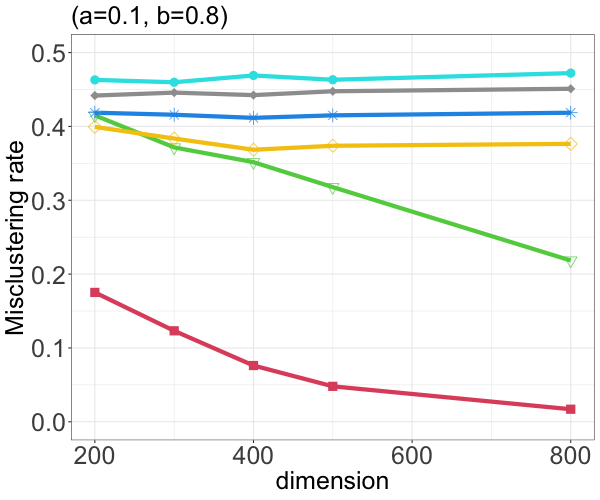}
\includegraphics[width=.32\textwidth]{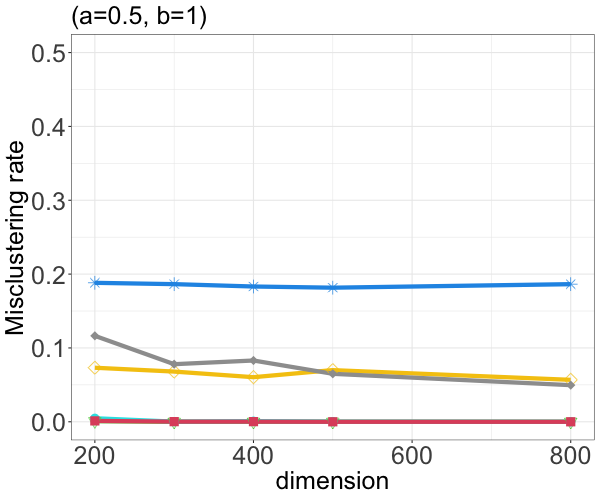}
\includegraphics[width=.32\textwidth]{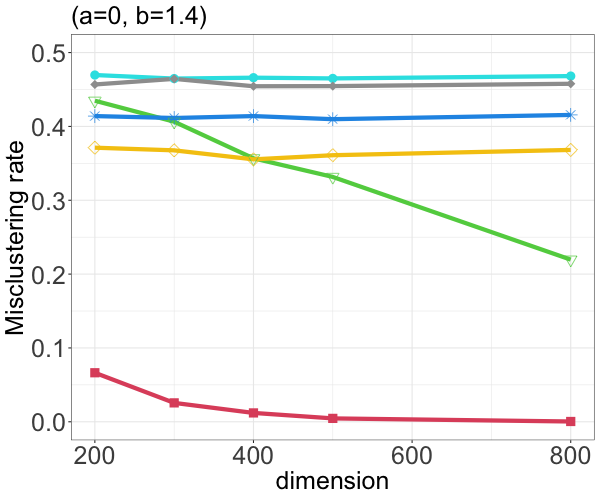}
\includegraphics[width=.32\textwidth]{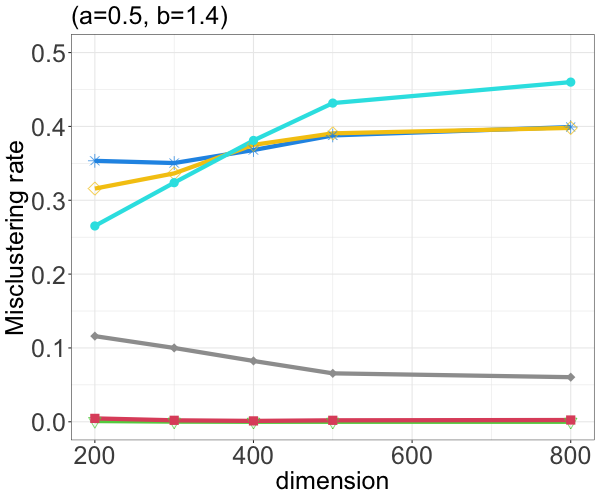}
\includegraphics[width=.7\textwidth]{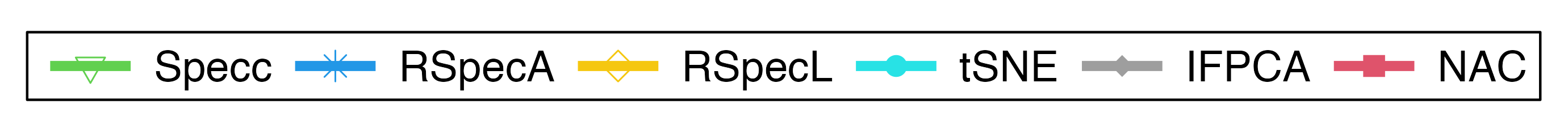}
\caption{Mis-clustering rates for Gaussian mixtures
$\mathcal N_d(0,\Sigma)$ and $\mathcal N_d(a\mathbf{1},b\Sigma)$ across dimensions.}
\label{fig:simulation-gaussian}
\end{figure}

We first consider Gaussian mixtures
$X_1\sim \mathcal N_d(0,\Sigma)$ and
$X_2\sim \mathcal N_d(a\mathbf 1,b\Sigma)$, where
$\Sigma_{ij}=0.1^{|i-j|}$. The parameters $a$ and $b$ control location and
scale differences, respectively. This setting directly tests the distinction
motivating NAC: location separation tends to produce symmetric within-neighbor
enrichment, whereas scale separation can produce asymmetric nearest-neighbor
structure.

Figure~\ref{fig:simulation-gaussian} reports mis-clustering rates across
dimensions. The columns correspond to location, scale, and combined
location-scale differences. In the location-separated settings, several methods
perform well when the signal is sufficiently strong, and NAC remains competitive.
In the scale-separated settings, many baselines remain close to random guessing,
while NAC improves rapidly with dimension. In the combined location-scale
settings, NAC retains strong performance because the criterion can use either
the symmetric statistic $Z_w$ or the asymmetric statistic $Z_d$, depending on
which signal is more prominent for the graph scale.

The six panels also highlight the different failure modes of the baselines. In
the weak location setting, all methods are challenged, reflecting the small
signal size. When the location signal is stronger, Specc and NAC both perform
well. In the scale-only settings, however, methods based primarily on symmetric
connectivity remain close to chance, while NAC improves as dimension increases.
This is consistent with the diagnostic in Section~\ref{sec:nn-asymmetry}: the
dispersed cluster need not form an internally dense nearest-neighbor subgraph,
so the contrast statistic becomes the more informative component. The combined
settings show that this additional component does not sacrifice performance
under location information; instead, the maximum criterion allows the stronger
signal to dominate.

\subsection{Non-Gaussian mixtures}
\label{sec:nongaussian}

\begin{figure}[!h]
\centering
\includegraphics[width=0.32\textwidth]{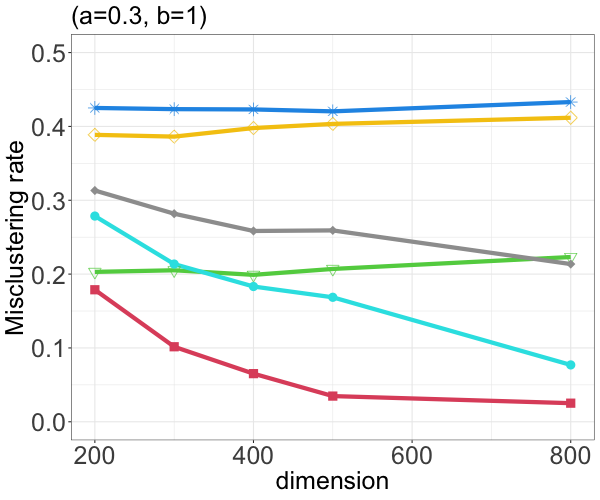}
\includegraphics[width=0.32\textwidth]{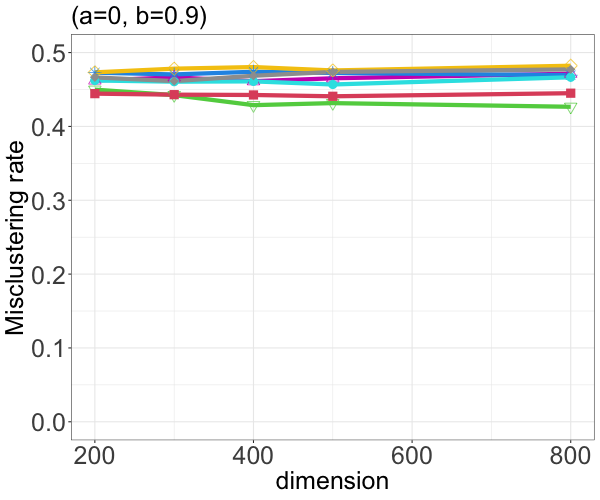}
\includegraphics[width=0.32\textwidth]{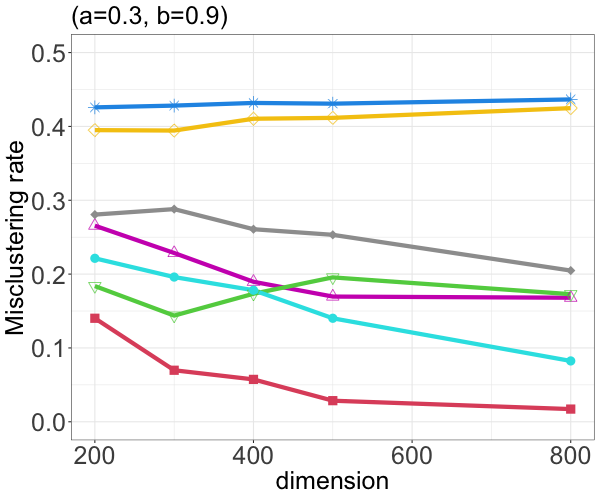}
\includegraphics[width=0.32\textwidth]{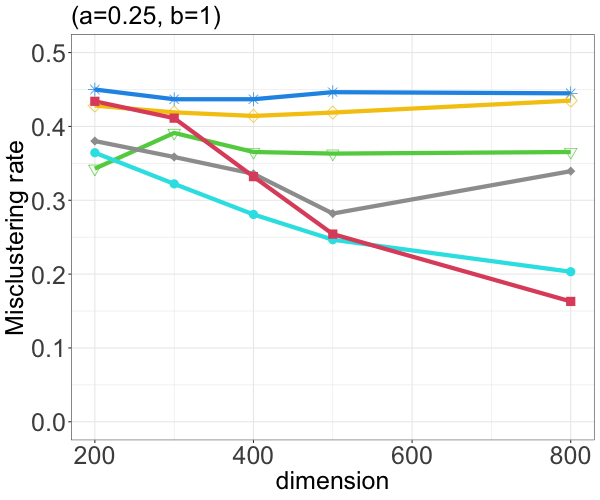}
\includegraphics[width=0.32\textwidth]{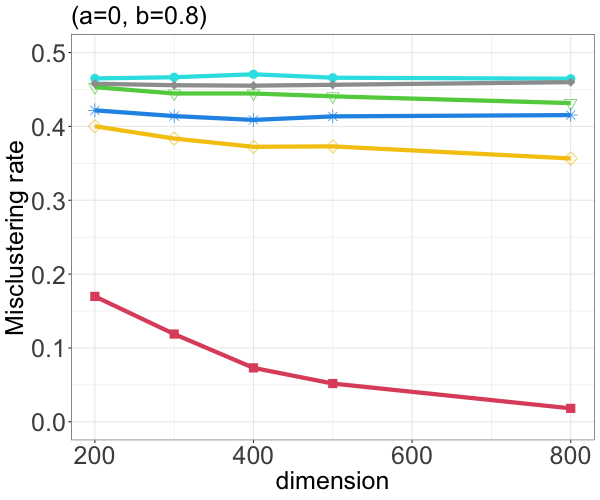}
\includegraphics[width=0.32\textwidth]{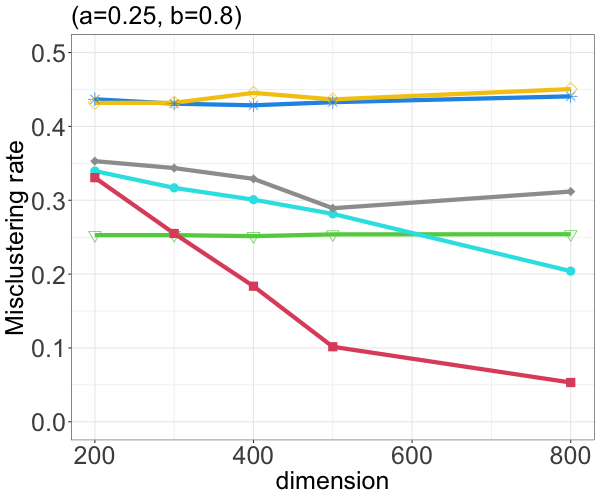}
\includegraphics[width=.7\textwidth]{shared_horizontal_legend.png}
\caption{Mis-clustering rates for multivariate $t$-distribution mixtures
$t_{20}(0,\Sigma)$ and $t_{20}(a\mathbf{1},b\Sigma)$. } 
\label{fig:simulation-t20} 
\end{figure}

We next consider heavy-tailed multivariate $t$-distribution mixtures to evaluate
NAC beyond the Gaussian settings above. Specifically, we generate two samples
from $t_{20}(0,\Sigma)$ and $t_{20}(a\mathbf 1,b\Sigma)$, where
$\Sigma_{ij}=0.1^{|i-j|}$. As in the Gaussian experiments, $a$ controls location
separation and $b$ controls scale separation. Figure~\ref{fig:simulation-t20}
reports representative results across dimensions.

The columns correspond to location, scale, and combined location-scale
differences. Under location separation, NAC remains competitive and improves as
the dimension increases. Under pure scale separation, Specc performs best in
these heavy-tailed settings, while NAC remains substantially more accurate than
most of the other baselines. Under combined location-scale differences, NAC
performs best, with a clear advantage that becomes more pronounced in higher
dimensions.

The pure-scale cases provide a useful qualification. For the milder scale alternative
($a=0, b=0.9$), all methods perform poorly, with mis-clustering rates remaining close
to chance. However, for the stronger scale alternative ($a=0, b=0.8$), NAC becomes
highly effective and its error decreases rapidly as the dimension increases. NAC also
performs very well in the pure-location setting ($a=0.3,b=1$) and in the combined
location-scale settings, often with a clear advantage in higher dimensions. Overall,
these results suggest that NAC is particularly effective when the distributional
difference is reflected in the nearest-neighbor structure, whether through location
separation, stronger scale changes, or a combination of the two.

\section{Gene-expression applications}
\label{sec:applications}

We apply NAC to four gene-expression datasets, each with two classes and more
features than observations
\citep{alon1999broad,golub1999molecular,gordon2002translation,
bhattacharjee2001classification}. These datasets represent the small-sample,
high-dimensional regime common in exploratory biomedical clustering, where the
form of separation between biological groups is not known in advance. The
datasets are summarized in Table~\ref{tab:dataset_info}. The algorithms are
applied without using the class labels; the labels are used only afterward to
compute mis-clustering rates.

\begin{table}[!h]
\centering
\caption{Gene-expression datasets used in the real-data analysis.}
\label{tab:dataset_info}
\setlength{\tabcolsep}{6pt}
\begin{tabular}{llcrr}
\toprule
Dataset & Abbrev. & Source & $n$ & $p$ \\
\midrule
Colon cancer & Cln & Alon et al. (1999) & 62 & 2,000 \\
Leukemia & Leuk & Golub et al. (1999) & 72 & 3,571 \\
Lung cancer (1) & Lung1 & Gordon et al. (2002) & 181 & 12,533 \\
Lung cancer (2) & Lung2 & Bhattacharjee et al. (2001) & 203 & 12,600 \\
\bottomrule
\end{tabular}
\end{table}

Table~\ref{tab:realdata-main} reports the mis-clustering rates. NAC achieves the best result on Colon cancer and ties for the best result on Lung cancer (2). On Leukemia, Specc performs best, while NAC remains close to the best result and outperforms several other baselines. On Lung cancer (1), IF-PCA performs best, suggesting that the separation in this dataset may be well captured by influential features and a low-dimensional principal-component representation. Overall, these results suggest that NAC is a useful complementary method for small-sample, high-dimensional clustering: across the four datasets, it achieves or matches the best performance on two datasets and remains competitive on the others.

\begin{table}[!h]
\centering
\caption{Mis-clustering rates on gene-expression datasets. Best in each row is
bolded.}
\label{tab:realdata-main}
\setlength{\tabcolsep}{4pt}
\begin{tabular}{lrrrrrr}
\toprule
Dataset & Specc & RSpec(A) & RSpec(L) & IF-PCA & $t$-SNE & NAC \\\midrule
Colon & 0.403 & 0.210 & 0.258 & 0.403 & 0.371 & \textbf{0.113} \\
Leukemia & \textbf{0.042} & 0.153 & 0.069 & 0.069 & 0.083 & 0.056 \\
Lung (1) & 0.099 & 0.254 & 0.403 & \textbf{0.033} & 0.326 & 0.116 \\
Lung (2) & \textbf{0.217} & 0.468 & 0.232 & \textbf{0.217} & 0.266 & \textbf{0.217} \\\bottomrule
\end{tabular}
\end{table}

\section{Conclusion}
\label{sec:discussion}

We proposed NAC, a nonparametric clustering method based on high-dimensional
nearest-neighbor asymmetry. By combining a standardized within-edge enrichment
statistic with a standardized asymmetry statistic, NAC adapts to both
location-like and scale-like separation without requiring the separation type to
be specified in advance. We use a fixed default $\kappa=1.55$ and select $k$ by
the largest optimized value of $M_{\kappa,k}$ over a grid of graph sizes; in
semi-supervised settings, labeled observations could instead be used to tune
$k$ and $\kappa$.


\bibliographystyle{plainnat}
\bibliography{neurips-ref}

\appendix

\appendix

\section{Proof of Theorem~\ref{thm:criterion}}
\label{app:proof}

We provide the algebraic argument behind Theorem~\ref{thm:criterion}. The proof
is written at the population level for a fixed directed $k$-nearest-neighbor
graph. The graph can be random through the data, but the comparison below
conditions on the limiting directed-neighbor probabilities. The goal is not to
prove global consistency of the nonconvex optimizer over all possible graphs,
but to explain why the proposed graph statistics select the true partition when
the nearest-neighbor probabilities exhibit the patterns described in
Section~\ref{sec:nn-asymmetry}.

Let the true cluster sizes be $m^*$ and $n^*$, and let $x^*$ denote the true
partition. Consider a candidate partition $x$ obtained by moving $\Delta_1$
observations from true cluster 1 to label 2 and $\Delta_2$ observations from
true cluster 2 to label 1. Thus
\[
m=m^*-\Delta_1+\Delta_2,\qquad
n=n^*-\Delta_2+\Delta_1.
\]
For fixed $k$, write
\[
p^*_{ab}=p_{ab,k}(x^*),\qquad a,b\in\{1,2\},
\]
for the directed-neighbor probability from true cluster $a$ to true cluster
$b$. Conditional on these graph-level probabilities, the expected within-label
edge counts under the candidate partition can be decomposed as
\begin{align}
\mathbb E_2 R_1(x)
=&\ p^*_{11} k m^*
\frac{(m^*-\Delta_1)(m^*-\Delta_1-1)}{m^*(m^*-1)}
+ p^*_{12} k m^*
\frac{(m^*-\Delta_1)\Delta_2}{m^*n^*} \notag\\
&\quad
+ p^*_{21} k n^*
\frac{(m^*-\Delta_1)\Delta_2}{m^*n^*}
+ p^*_{22} k n^*
\frac{\Delta_2(\Delta_2-1)}{n^*(n^*-1)},
\label{eq:ER1-proof}
\end{align}
and
\begin{align}
\mathbb E_2 R_2(x)
=&\ p^*_{22} k n^*
\frac{(n^*-\Delta_2)(n^*-\Delta_2-1)}{n^*(n^*-1)}
+ p^*_{21} k n^*
\frac{(n^*-\Delta_2)\Delta_1}{n^*m^*} \notag\\
&\quad
+ p^*_{12} k m^*
\frac{(n^*-\Delta_2)\Delta_1}{n^*m^*}
+ p^*_{11} k m^*
\frac{\Delta_1(\Delta_1-1)}{m^*(m^*-1)}.
\label{eq:ER2-proof}
\end{align}
Here $\mathbb E_2$ denotes expectation under random relabeling within the four
groups induced by the true and candidate partitions. The four terms in
\eqref{eq:ER1-proof} correspond to edges from true cluster 1 to true cluster 1
that remain inside candidate label 1, edges from true cluster 1 to true cluster
2 that fall inside candidate label 1, edges from true cluster 2 to true cluster
1 that fall inside candidate label 1, and edges from true cluster 2 to true
cluster 2 that fall inside candidate label 1. Equation~\eqref{eq:ER2-proof} has
the analogous interpretation for candidate label 2.

We first study the difference statistic, since its algebra is simpler. From
\eqref{eq:ER1-proof}--\eqref{eq:ER2-proof},
\begin{align}
\mathbb E_2 R_d(x)
=&\ \mathbb E_2 R_1(x)-\mathbb E_2 R_2(x) \notag\\
=&\ p^*_{11}km^* - p^*_{22}kn^*
- k\Delta_1\left(p^*_{11}-\frac{n^*}{m^*}p^*_{22}
+\frac{N}{m^*}\right)
+ k\Delta_2\left(p^*_{22}-\frac{m^*}{n^*}p^*_{11}
+\frac{N}{n^*}\right).
\label{eq:ERd-proof}
\end{align}
Since $\mu_d(x)=k(m-n)$ and
$m-n=m^*-n^*-2\Delta_1+2\Delta_2$, we have
\begin{align}
\mathbb E_2 R_d(x)-\mu_d(x)
=&\ k\{p^*_{11}m^* - p^*_{22}n^*-(m^*-n^*)\} \notag\\
&\quad
- k\Delta_1\left(p^*_{11}-\frac{n^*}{m^*}p^*_{22}
+\frac{N}{m^*}-2\right)
+ k\Delta_2\left(p^*_{22}-\frac{m^*}{n^*}p^*_{11}
+\frac{N}{n^*}-2\right) \notag\\
=&\ k\{p^*_{11}m^* - p^*_{22}n^*-(m^*-n^*)\}
\left(1-\frac{\Delta_1}{m^*}-\frac{\Delta_2}{n^*}\right).
\label{eq:ERd-centered-proof}
\end{align}
The denominator of the standardized statistic $Z_d$ is proportional, up to a
positive graph-dependent constant, to
\[
\sqrt{(m^*-\Delta_1+\Delta_2)(n^*+\Delta_1-\Delta_2)}.
\]
Therefore the signed standardized contrast can be written as
\[
\widetilde Z_d(x)
=
\frac{\mathbb E_2 R_d(x)-\mu_d(x)}{\sigma_d(x)}
=
k\{p^*_{11}m^* - p^*_{22}n^*-(m^*-n^*)\}
f_d(\Delta_1,\Delta_2),
\]
up to a positive constant independent of $\Delta_1,\Delta_2$, where
\[
f_d(\Delta_1,\Delta_2)
=
\frac{
1-\Delta_1/m^*-\Delta_2/n^*
}{
\sqrt{(m^*-\Delta_1+\Delta_2)(n^*+\Delta_1-\Delta_2)}
}.
\]
A direct calculation gives
\[
\frac{\partial f_d}{\partial \Delta_1}
=
-\frac{N}{2m^*n^*}
\frac{
(n^*-\Delta_2)(m^*-\Delta_1+2\Delta_2)+\Delta_1\Delta_2
}{
(m^*-\Delta_1+\Delta_2)^{3/2}
(n^*+\Delta_1-\Delta_2)^{3/2}
}<0,
\]
and
\[
\frac{\partial f_d}{\partial \Delta_2}
=
-\frac{N}{2m^*n^*}
\frac{
(m^*-\Delta_1)(n^*-\Delta_2+2\Delta_1)+\Delta_1\Delta_2
}{
(m^*-\Delta_1+\Delta_2)^{3/2}
(n^*+\Delta_1-\Delta_2)^{3/2}
}<0.
\]
Thus $f_d$ is maximized at $\Delta_1=\Delta_2=0$, with value
$1/\sqrt{m^*n^*}$, and minimized at
$\Delta_1=m^*,\Delta_2=n^*$, with value $-1/\sqrt{m^*n^*}$.

Under Pattern 2,
\[
p^*_{11}>\frac{m^*-1}{N-1},
\qquad
p^*_{22}<\frac{n^*-1}{N-1},
\]
so
\begin{align*}
p^*_{11}m^*-p^*_{22}n^*-(m^*-n^*)
&>
\frac{m^*(m^*-1)}{N-1}
-\frac{n^*(n^*-1)}{N-1}
-(m^*-n^*) \\
&=0.
\end{align*}
Hence the signed contrast $\widetilde Z_d$ is maximized at
$\Delta_1=\Delta_2=0$, i.e., at $x=x^*$. Under Pattern 3, the same coefficient
is negative, so the signed contrast is maximized at
$\Delta_1=m^*,\Delta_2=n^*$, i.e., at the switched labeling $x=1-x^*$. Since the
main text defines
\[
Z_d(x)=|\widetilde Z_d(x)|,
\]
the absolute standardized contrast is maximized by the true partition up to
label switching in both asymmetric patterns.

We next study the weighted statistic. Let
\[
A_w
=
m^*(n^*-1)p^*_{11}
+n^*(m^*-1)p^*_{22}
-\frac{(m^*-1)(n^*-1)N}{N-1}.
\]
Substituting \eqref{eq:ER1-proof}--\eqref{eq:ER2-proof} into
\[
R_w(x)=\frac{n-1}{N-2}R_1(x)+\frac{m-1}{N-2}R_2(x)
\]
and subtracting the permutation-null centering term gives
\begin{align}
\mathbb E_2 R_w(x)-\mu_w(x)
=
\frac{kA_w}{N-2}
\Bigg\{
&1
+\frac{\Delta_1^2}{m^*(m^*-1)}
+\frac{\Delta_2^2}{n^*(n^*-1)}
+\frac{2\Delta_1\Delta_2}{m^*n^*} \notag\\
&-\frac{(2m^*-1)\Delta_1}{m^*(m^*-1)}
-\frac{(2n^*-1)\Delta_2}{n^*(n^*-1)}
\Bigg\}.
\label{eq:ERw-centered-proof}
\end{align}
The denominator of $Z_w$ is proportional, up to a positive graph-dependent
constant, to
\[
\sqrt{
(m^*-\Delta_1+\Delta_2)
(n^*+\Delta_1-\Delta_2)
(m^*-\Delta_1+\Delta_2-1)
(n^*+\Delta_1-\Delta_2-1)
}.
\]
Thus
\[
\mathbb E_2 Z_w(x)
=
\frac{kA_w}{N-2} f_w(\Delta_1,\Delta_2),
\]
again up to a positive constant independent of $\Delta_1,\Delta_2$, where
\begin{align}
f_w(\Delta_1,\Delta_2)
=
\frac{
1
+\frac{\Delta_1^2}{m^*(m^*-1)}
+\frac{\Delta_2^2}{n^*(n^*-1)}
+\frac{2\Delta_1\Delta_2}{m^*n^*}
-\frac{(2m^*-1)\Delta_1}{m^*(m^*-1)}
-\frac{(2n^*-1)\Delta_2}{n^*(n^*-1)}
}{
\sqrt{
(m^*-\Delta_1+\Delta_2)
(n^*+\Delta_1-\Delta_2)
(m^*-\Delta_1+\Delta_2-1)
(n^*+\Delta_1-\Delta_2-1)
}
}.
\label{eq:fw-proof}
\end{align}

It remains to identify where $f_w$ is maximized. Define
\[
\widetilde f_d(\Delta_1,\Delta_2)
=
\frac{
1-\Delta_1/(m^*-1)-\Delta_2/(n^*-1)
}{
\sqrt{
(m^*-\Delta_1+\Delta_2-1)
(n^*+\Delta_1-\Delta_2-1)
}
}.
\]
By the same monotonicity argument used for $f_d$, for
$\Delta_1\in[0,m^*-1]$ and $\Delta_2\in[0,n^*-1]$,
$\widetilde f_d$ is maximized at $\Delta_1=\Delta_2=0$ and minimized at
$\Delta_1=m^*-1,\Delta_2=n^*-1$. Moreover,
\[
f_w(\Delta_1,\Delta_2)
=
f_d(\Delta_1,\Delta_2)\widetilde f_d(\Delta_1,\Delta_2)
-
h(\Delta_1,\Delta_2),
\]
where
\[
h(\Delta_1,\Delta_2)
=
\frac{
\frac{N-2}{m^*n^*(m^*-1)(n^*-1)}\Delta_1\Delta_2
}{
\sqrt{
(m^*-\Delta_1+\Delta_2)
(n^*+\Delta_1-\Delta_2)
(m^*-\Delta_1+\Delta_2-1)
(n^*+\Delta_1-\Delta_2-1)
}
}.
\]
Since $h(\Delta_1,\Delta_2)>0$ whenever $\Delta_1\Delta_2>0$, the product
$f_d\widetilde f_d$ is largest at the perfectly matched or perfectly switched
labelings, and the subtraction term rules out interior simultaneous
mislabelings from exceeding the perfectly matched value.

The remaining corner cases are
\[
(\Delta_1,\Delta_2)=(m^*,n^*),\quad
(m^*,n^*-1),\quad
(m^*-1,n^*).
\]
They satisfy
\[
f_w(m^*,n^*)
=
\frac{1}{\sqrt{m^*n^*(m^*-1)(n^*-1)}}
=
f_w(0,0),
\]
while
\[
f_w(m^*,n^*-1)
=
\frac{
\sqrt{\left(1-\frac{2}{n^*}\right)
\left(1-\frac{2}{m^*+1}\right)}
}{
\sqrt{m^*n^*(m^*-1)(n^*-1)}
}
<f_w(0,0),
\]
and
\[
f_w(m^*-1,n^*)
=
\frac{
\sqrt{\left(1-\frac{2}{m^*}\right)
\left(1-\frac{2}{n^*+1}\right)}
}{
\sqrt{m^*n^*(m^*-1)(n^*-1)}
}
<f_w(0,0).
\]
Therefore $f_w(\Delta_1,\Delta_2)$ is maximized at
$\Delta_1=\Delta_2=0$ or at $\Delta_1=m^*,\Delta_2=n^*$, corresponding to
$x=x^*$ and the switched labeling $x=1-x^*$.

Under Pattern 1,
\[
p^*_{11}>\frac{m^*-1}{N-1},
\qquad
p^*_{22}>\frac{n^*-1}{N-1},
\]
so
\begin{align*}
A_w
&=
m^*(n^*-1)p^*_{11}
+n^*(m^*-1)p^*_{22}
-\frac{(m^*-1)(n^*-1)N}{N-1} \\
&>
m^*(n^*-1)\frac{m^*-1}{N-1}
+n^*(m^*-1)\frac{n^*-1}{N-1}
-\frac{(m^*-1)(n^*-1)N}{N-1}
=0.
\end{align*}
Thus $\mathbb E_2 Z_w(x)$ is maximized at
$\Delta_1=\Delta_2=0$ or $\Delta_1=m^*,\Delta_2=n^*$, that is, at the true
partition up to label switching.

Combining the arguments above proves the theorem.

\section{Optional faster graph-size search}
\label{app:optimization}

The main experiments evaluate a grid of graph sizes and select the partition
corresponding to the largest optimized value of $M_{\kappa,k}$, as described in
Section~\ref{sec:optimization}. When evaluating a
large grid of $k$ values is expensive, a faster graph-size search can be used as
an optimization shortcut. Rather than applying this shortcut directly to
$M_{\kappa,k}$, which need not be unimodal as a function of $k$, we apply it to
the two optimized components separately.

Let
\[
F_w(k)=\max_x Z_w^{(k)}(x),
\qquad
F_d(k)=\max_x Z_d^{(k)}(x),
\]
where the superscript indicates that the statistic is computed on the
$k$-nearest-neighbor graph. For a fixed $\kappa$,
\[
\max_k \max_x M_{\kappa,k}(x)
=
\max\left\{\max_k F_w(k),\ \kappa\max_k F_d(k)\right\}.
\]
Thus, if the relevant optimized component is approximately unimodal over $k$, a
ternary-search refinement can be applied to $F_w$ and $F_d$ separately.

\begin{algorithm}[h]
\caption{Ternary search over graph size}
\label{alg:ternary-k}
\begin{algorithmic}[1]
\Require Objective value $F(k)$ for the $k$-nearest-neighbor graph.
\Ensure Approximate maximizing graph size.
\State Set $l=1$ and $r=N-1$.
\While{$r-l>1$}
    \State Set $k_1=\lfloor l+(r-l)/3\rfloor$ and
    $k_2=\lfloor r-(r-l)/3\rfloor$.
    \State Evaluate $F(k_1)$ and $F(k_2)$.
    \If{$F(k_1)<F(k_2)$}
        \State Set $l=k_1$.
    \Else
        \State Set $r=k_2$.
    \EndIf
\EndWhile
\State \Return the graph size with the largest evaluated objective value.
\end{algorithmic}
\end{algorithm}

In practice, Algorithm~\ref{alg:ternary-k} can be run once with
$F=F_w$ and once with $F=F_d$. The resulting graph sizes and partitions are then
treated as candidate solutions and compared using the full criterion
$M_{\kappa,k}$. It is intended only as a computational shortcut; the main experiments use the
more robust grid-based procedure that directly evaluates the candidate graph
sizes.

\section{Calibration and interpretation of $\kappa$}
\label{app:kappa}

The criterion
\[
M_{\kappa,k}(x)=\max\{Z_w(x),\kappa Z_d(x)\}
\]
uses $\kappa$ to balance the weighted statistic $Z_w$ and the difference
statistic $Z_d$. Although both statistics are standardized under the
random-labeling null, their optimized values are not directly interchangeable
after maximization over candidate partitions. We therefore use a fixed default
value of $\kappa$, chosen from representative location and scale settings and
held fixed across all experiments.

\begin{figure}[!htp]
\includegraphics[width=0.49\textwidth]{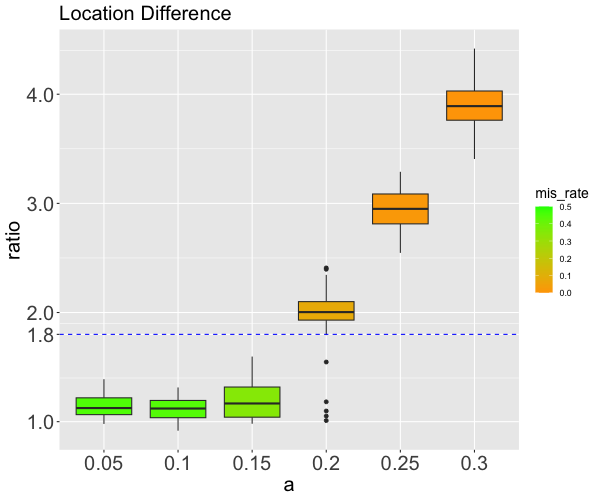}
\includegraphics[width=0.49\textwidth]{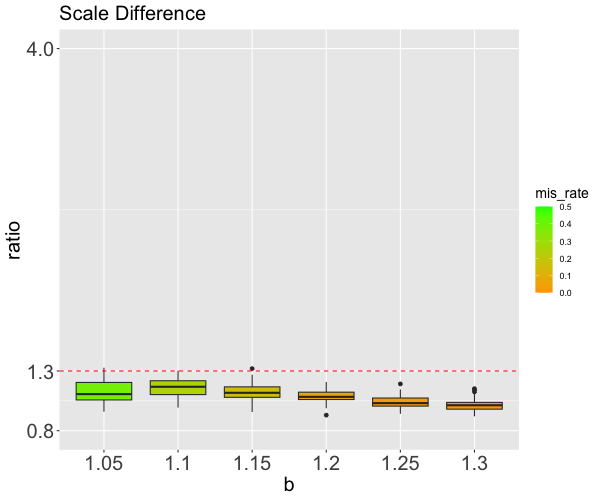}
\caption{Calibration of $\kappa$ using the ratio $r=Z_w/Z_d$. Left: location
difference with fixed scale. Right: scale difference with fixed location. The
reference values $r=1.8$ and $r=1.3$ bracket a stable range for balancing the
two statistics; we use the midpoint $\kappa=1.55$ in all reported experiments.} 
\label{fig:kappa-calibration}
\end{figure}

To calibrate $\kappa$, we examine the ratio
\[
r=\frac{Z_w}{Z_d}
\]
at high-scoring partitions in settings where the relevant signal is known.
In location-difference settings, $Z_w$ should be favored; in scale-difference
settings, $Z_d$ should be favored. Figure~\ref{fig:kappa-calibration} plots this
ratio as the signal strength varies. In the location panel, the lower end of the
high-performing $Z_w$ region is around $r=1.8$. In the scale panel, the upper end
of the high-performing $Z_d$ region is around $r=1.3$. We therefore choose the
midpoint,
\[
\kappa = \frac{1.8+1.3}{2}=1.55,
\]
as the default value.

This calibration is not performed separately for each simulation setting or real
dataset. It is used only to choose a single default. If prior scientific
information suggests that a problem is predominantly scale-driven, a larger
weight on $Z_d$ can be considered as a sensitivity analysis; if the problem is
believed to be purely location-driven, a smaller value can be considered. The
main empirical results use the same $\kappa=1.55$ throughout.

\section{Choice of \texttt{kpar} in \texttt{specc}}
\label{app:specc-kpar}

For \texttt{specc}, the kernel-width parameter \texttt{kpar} can affect
performance substantially. In high-dimensional settings, we found that the
default \texttt{kpar="automatic"} option can be unstable and may produce
unexpected errors, especially when the dimension is large. The alternative
\texttt{kpar="local"} option uses a local heuristic for the kernel width and is
more reliable in the settings considered here.

\begin{figure}[h]
\centering
\includegraphics[width=\textwidth]{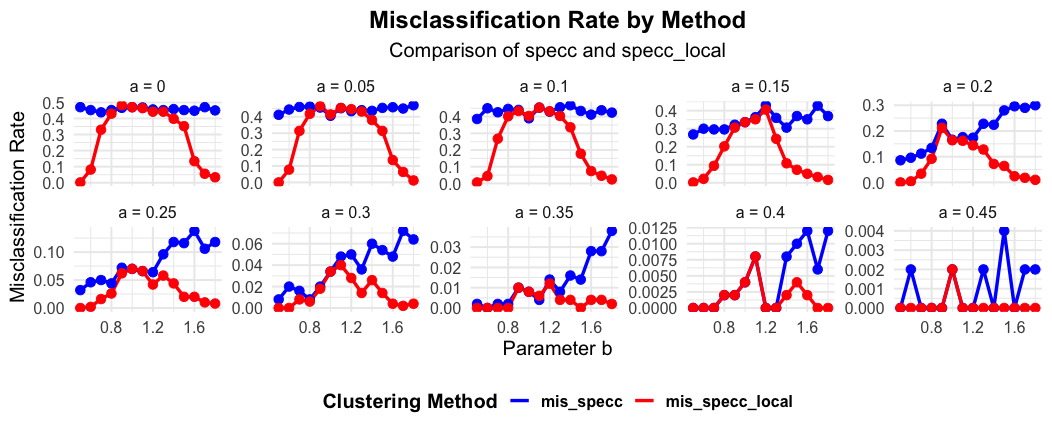}
\caption{Comparison of two \texttt{kpar} choices in \texttt{specc} for
high-dimensional Gaussian mixtures: \texttt{kpar="automatic"} and
\texttt{kpar="local"}.}
\label{fig:specc-kpar}
\end{figure}

To compare the two choices in a high-dimensional setting, we generated two
samples from $\mathcal N_d(0,\Sigma)$ and $\mathcal N_d(a\mathbf 1,b\Sigma)$
with $d=300$, $\Sigma_{ij}=0.1^{|i-j|}$, and 50 observations in each cluster.
Figure~\ref{fig:specc-kpar} compares the default automatic choice with the local
choice across a range of location and scale parameters. The local choice
generally performs as well as or better than the automatic choice and avoids the
instability observed with the default option. We therefore use
\texttt{kpar="local"} when applying \texttt{specc} to high-dimensional data.

\end{document}